\documentclass[prl,
twocolumn,
showpacs,aps,amssymb,amsfonts,amsmath
]{revtex4}
\usepackage[vcentermath]{youngtab}
\usepackage{amssymb}
\usepackage{graphicx}

\begin{document}
\title{\bf The Pauli principle and magnetism}
\author{Alexander A. Klyachko}
\email{klyachko@fen.bilkent.edu.tr}

\affiliation{ Bilkent University, Faculty of Science, 06800,
Bilkent, Ankara, Turkey}

\begin{abstract}
According to Heisenberg, ferromagnetism stems from
the Pauli magic mediating
between Coulomb interaction and electrons' spins. The primary aim of
this paper is to turn the
magic into an algebra by setting a precise bound to  the degree
a given electron distribution can affect spin. An
application of the
resulting {\it spin-orbital Pauli constraints\/} to Fe, Co, and Ni
provides a new insight into the origin of
magnetic moment in these archetypical ferromagnets.
\end{abstract}

\maketitle
\paragraph{\bf Introduction.} Everyone agrees that the initial
form of the Pauli principle -- {\it no state can be occupied by
more than one electron} -- is obsolete and must be replaced by the
concept of skew-symmetry of a multi-electron  wave functions.
Nevertheless,
in treatment of Fermi-Dirac statistics
we tacitely
 assume  that the Pauli condition $\langle\psi|\rho|\psi\rangle\le 1$
is the only constraint on the fermionic density matrix $\rho$.
In fact, there are many more of them, generically
given by linear inequalities on eigenvalues of $\rho$ \cite{B-D,Ruskai, Klyachko08}.
For finite dimensional one-electron space $\mathcal{H}$,
at least
in principle,  all Pauli constraints can be found
\cite{Klyachko08}.

The one electron space $\mathcal{H}=\mathcal{H}_\ell\otimes \mathcal{H}_s$ splits
into
orbital  and spin degrees of freedom.  Here we assume that   $N$-electron state $\Psi\in \wedge^N\mathcal{H}$ has a well defined total spin $S$ and the respective spin density matrix $\rho_S$ along with the orbital one $\rho_\ell$.
Let $\mu_1\ge\mu_2\ge\mu_3
\ge\cdots$ and $\nu_1\ge\nu_2\ge\nu_3\ge\ldots$ be spectra of spin and orbital density matrices normalized to the the unit probability  $\sum_j\mu_j=1$ and to the number of particles  $\sum_i\nu_i=N$ respectively.
In this setting the Pauli constraints amount to a system of linear inequalities on  orbital and spin
occupation numbers $\nu_i$, $\mu_j$ \cite{Klyachko08}. They  will be referred  below as
{\it spin-orbital Pauli constraints\/.}
As a toy example, consider the Pauli constraints for  $d^7$ shell in low spin  configuration $S=\frac12$ borrowed from \cite
{Klyachko08}
\begin{equation}\label{low_spin1}
\small\begin{array}{l}
M\le 3-2(\nu_3-\nu_4);\quad M\ge 2(\nu_3-\nu_5)-3;\\
M\le 3-2(\nu_4-\nu_5);\quad M\ge 1-2(\nu_2-\nu_4+2\nu_5);\\
\hspace{2cm}
\nu_3+\nu_4-\nu_5\le 3.
\end{array}
\end{equation}
 Here $M=\mu_1-\mu_2$ is the maximal possible spin magnetic moment for a given spin density $\mu$ and I skip the initial Pauli constraint $\nu\le2$. In a BCC crystal favouring  $t_{2g}$ orbitals $\nu_1=\nu_2=\nu_3=a\ge b=\nu_4=\nu_5$ we get the bound $M\le 10-5a$ that falls  short
of the nominal
spin-$\frac12$
value $M=1\mu_B$ for $a>\frac95$. 
It should be emphasized that the spin moment reduction
came directly from the Pauli kinematics
with no interaction involved.

I will give more realistic examples below and for now
put this insight into a historical context.
First of all, it sheds light on
80 years old puzzle of reduction atomic spin moments in ferromagnets.
The problem was raised by Pauli  at Solvay congress  on magnetism in 1930 \cite{Pauli30}.
Being unable to resolve the issue on intra-atomic level he reluctantly  conceded that ``{\it it seems rather that several atoms should participate}"
and 
inscribed the riddle  into the list of three open problems in
magnetism. 
Later on  Stoner has turned the unsolved Pauli problem into a paradigm  of itinerate magnetism

{\footnotesize
\begin{quote} All the earlier atomic theories of
ferromagnetism are confronted with the difficulty that the
saturation magnetization does not correspond to an integral number
of Bohr magnetons per atom, a difficulty which could be met only
by an artificial and ad hoc conception of a metal as containing an
equilibrium mixture of neutral atoms and ions of different
moments, including zero. \cite[p. 69]{Stoner47}
\end{quote}}
The very existence of the spin-orbital Pauli constraints
makes the Stoner's argument void
and relegates it to history books along with its numerous
variations repeated in almost every modern text on ferromagnetism

{\footnotesize\begin{quote}
 [T]he experimental magnetic moment $\mu=2.2\,\mu_B$
 is not an integer number, which indicates presence of some
 fraction of itinerant electrons. \cite{Katanin10}

The magnetism of the iron-series transition-metal elements is
caused by extended delocalized or itinerate electrons.
The itinerate character is epitomized by noninteger spin moments
per atom [\ldots].
\cite[p. 52]{Skomski08}

 Today everyone agrees that some degree of itinerancy must be
granted to the $d$-electrons in order to explain many basic
experimental facts, such as the saturate nonintegral number of
Bohr magnetons per atom. \cite{Mota86}
\end{quote}}

The idea  that the reduction effect comes directly from the Pauli kinematics,
as opposed to
the ambiguous
concept of
exchange interaction, 
goes back to Feynman \footnote{``Soon
after the development of quantum mechanics, it was noticed that
there is a very strong {\it apparent\,} force -- not a magnetic
force or any other kind of actual force, but only an apparent
force -- trying to line the spins of nearby electrons {\it
opposite} to one another.
Sometimes this spin-turning force is called the {\it exchange
force}, but that only makes it more mysterious -- it is not a very
good term. It is just because of the exclusion principle that
electrons have a tendency to make their spins opposite." Feynman's
Lectures in Physics, 37-2.}.
 He, however, could not fulfill the prophesy since the relevant math was not available at that time.


The current  study rests upon
a calculation of the Pauli constraints
for all high spin $d$-shell configurations.
Most of them, whether to good or to bad,
turn  out to be counterintuitive and
defy any simple explanation. Yet that very
incomprehensibility makes the Pauli constraints indispensable for
an expansion of our insight beyond the horizon of
intuition into
a new unexplored realm hidden beneath the surface of common sense.
I will discuss them
at some length in the next section.



\begin{figure}[t]
\includegraphics[width=6cm]{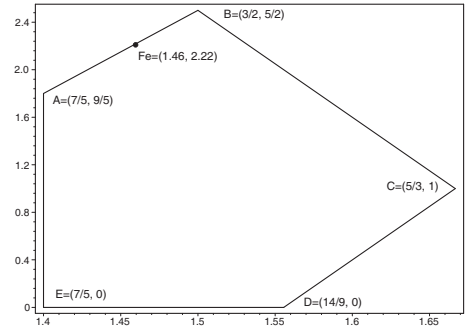}
\caption{ \label{fig1} The Pauli constraints on spin magnetic moment
for $d^7$ configuration in BCC symmetry versus the occupation number
$a$ of a $t_{2g}$ orbital. All points within the pentagon $ABCDE$
are admissible. The black dot represents iron's data.}
\end{figure}

The rest of the paper is focused on various  manifestations of the Pauli constraints in ferromagnetic iron, 
cobalt, and nickel.
Densities of $d$-electrons in these archetypical ferromagnets
were accurately measured by Jauch\&Reehuis \cite{Jauch07, Jauch08, Jauch09}.
Plugging these data into spin-orbital constraints yields a  {\it Pauli upper bound} on the spin magnetic moment.
For iron
it coincides with the actual
value $2.22\;\mu_B$ of the magnetic moment, see Fig~\ref{fig1}. Thus the Pauli constraints provide an
accurate  account for the observed atomic magnetic moment.
A similar saturation of the Pauli bound holds in cobalt,
but not in nickel where the spin magnetic  moment is far below 
of that bound.

The last section deals with
an impact of
the Pauli constraints on quantum statistics.
It can be best  seen in {\it crossovers\/} between different regimes  in magnetisation curve of iron. 
Indeed, at zero temperature  the atomic state in iron saturates {\it three\/} independent Pauli constraints, while at high temperature the
Pauli constraints become  irrelevant. As the temperature decreases, the spin state saturate the Pauli constraints one by one that effectively quenches  certain degrees of freedom and qualitatively changes the thermal evolution.
 A calculation predicts  two such crossovers  roughly at about $460^\circ$K and $180^\circ$K. The first one has been previously reported by K\"obler \cite{Koebler03} at $467^\circ$K,  while the second one
was buried unnoticed  in Pauthenet's
 extremely accurate
study   of high field susceptibility \cite{Pauthenet82},
which exhibits a very clean   quantum crossover at $200^\circ$K shown in Fig~\ref{fig4}. See also recent book \cite{Koebler09} on the magnetic crossovers phenomenology.

\paragraph{\bf A first glimpse into the Pauli constraints.}
Consider a spin polarized system
with orbital
occupancies $\nu_1\ge\nu_2\ge\nu_3\cdots$.
In the simplest case of
{\it two\/} spin-polarized electrons the nonzero eigenvalues
are evenly degenerated $\nu_{2i-1}=\nu_{2i}$
that gives
the simplest  constraint beyond the initial Pauli condition $\nu_i\le 1$.
In particular, a spin-polarized $d^2$-shell must have
$\nu=(\nu_1,\nu_1,\nu_3,\nu_3,0)$.
By a similar reason orbital occupancies
of spin polarized $d^3,d^7,d^8$ shells
must be
of the form $(1,\nu_2,\nu_2,\nu_4,\nu_4)$,
$(\nu_1,\nu_1,\nu_3,\nu_3,1)$, $(2,\nu_2,\nu_2,\nu_4,\nu_4)$
respectively \cite{Klyachko08}.
In particular, sphericall symmetry $\nu=(a,a,a,a,a)$
is incompatible with full spin polarization 
, i.e. either
the charge
must be deformed  or the atomic  spin moment reduced.
When the number of electrons and dimension of the shell increase,
the extended Pauli constraints turn into inequalities rather than
equations. In that respect they are getting closer to the initial
Pauli principle.
For three spin-polarized electrons in $f$-shell
they are as follows
$$
\begin{array}{rr}\nu_2+\nu_3+\nu_4+\nu_5\le
2;&\qquad\nu_1+\nu_3+\nu_4+\nu_6\le 2;\cr
\nu_1+\nu_2+\nu_4+\nu_7\le 2;&\qquad\nu_1+\nu_2+\nu_5+\nu_6\le 2.
\end{array}$$
These constraints hold for any separable spin-orbital state and a failure of any of them
is a signature  of spin-orbital entanglement.
All spinless Pauli constraints for shells of dimension $\le10$
can be found in \cite{Klyachko08}.
Beyond that range
only
few of them are known, like
 the following ones for {\it three\/} elelctron shell
 of even dimension $n$
$$\nu_i+\nu_j\le 1,\quad i+j=n+1.$$
They clearly
supersede the initial Pauli principle $\nu_i\le1$.


The purely orbital Pauli constraints allow only to detect {\it an onset\/} of the spin-orbital entanglement and the resulting  spin moment reduction 
with no estimation of its magnitude. To move forward, we have to
include into consideration
spin density matrix $\rho_s$.
 Joint constraints on
both orbital $\nu$ and spin $\mu$ occupancies
have been treated theoretically in \cite{Klyachko08}. The actual calculation is rather involved. The results are summarised  in two tables below.

\begin{center}
\scriptsize{
\begin{tabular}[t]{|c|c|}
\hline
\multicolumn{2}{|c|}{\sc Spin-orbital Pauli constraints for high spin $d^7$ shell}  \\
\hline
$[0,0,1,-1,-2 \mid 0,1,0,2]\leq-1$ & $[1,0,0,1,-1\mid-1,0,0,1]\leq 2$\\
$[0,1,0,-1,-2\mid0,0,1,2]\leq -1$ & $[0,1,1,0,-1\mid-1,0,0,1]\leq 2$\\
$[0,1,-1,-2,0\mid0,2,0,1]\leq -1$ & $[0,0,1,-1,1\mid1,-1,0,0]\leq 2$\\
$[0,1,0,-2,-1\mid0,1,0,2]\leq -1$ & $[0,1,1,-1,0\mid0,-1,0,1]\leq 2$\\
$[0,1,-2,-1,0\mid0,2,1,0]\leq -1$ & $[1,-1,0,0,1\mid1,-1,0,0]\leq 2$\\
$[1,-2,-1,0,0\mid1,2,0,0]\leq -1$ & $[1,-1,0,0,1\mid0,1,-1,0]\leq 2$\\
$[1,-1,0,0,-2\mid0,1,0,1]\leq -1$ & $[1,0,1,-1,0\mid-1,0,0,1]\leq 2$\\
$[1,0,-1,0,-2\mid0,0,1,2]\leq -1$ & $[0,1,-1,0,1\mid0,1,-1,0]\leq 2$\\\cline{2-2}
$[1,-1,0,-2,0\mid0,1,2,0]\leq -1$ & $[0,0,1,-1,0\mid1,0,0,0]\leq 1$\\
$[1,0,0,-2,-1\mid0,0,1,2]\leq -1$ & $[1,-1,0,0,0\mid1,0,0,0]\leq 1$\\
$[1,0,-2,-1,0\mid0,1,2,0]\leq -1$ & $[1,0,-1,0,0\mid0,1,0,0]\leq 1$\\
$[1,0,-2,-1,0\mid0,2,0,1]\leq -1$ & $[1,0,0,-1,0\mid0,0,1,0]\leq 1$\\
$[0,-1,1,-2,0\mid2,0,0,1]\leq -1$ & $[1,0,0,0,-1\mid0,0,0,1]\leq 1$\\
$[0,0,1,-2,-1\mid1,0,0,2]\leq -1$ & $[0,1,0,-1,0\mid0,1,0,0]\leq 1$\\\cline{2-2}
$[-1,0,1,-2,0\mid2,0,1,0]\leq -1$ & $[0,0,1,-1,-1\mid0,0,1,2]\leq 0$\\
$[1,-2,-1,0,0\mid2,0,1,0]\leq -1$ & $[0,-1,0,-1,1\mid2,0,0,1]\leq 0$\\
$[1,-2,0,-1,0\mid2,0,0,1]\leq -1$ & $[0,-1,-1,0,1\mid1,2,0,0]\leq 0$\\\cline{1-1}
$[0,1,1,0,0\mid0,0,-1,0]\leq 3$ & $[-1,0,0,-1,1\mid2,0,1,0]\leq 0$\\
$[1,1,0,0,0\mid-1,0,0,0]\leq 3$ & $[0,-1,-1,0,1\mid2,0,1,0]\leq 0$\\
$[1,0,1,0,0\mid0,-1,0,0]\leq 3$ & $[-1,0,-1,0,1\mid2,1,0,0]\leq 0$\\\cline{2-2}
$[1,0,0,1,0\mid0,0,-1,0]\leq 3$ & $[1,0,1,0,1\mid1,0,0,0]\leq 5$\\
$[1,0,0,0,1\mid0,0,0,-1]\leq 3$ & $[1,1,0,0,1\mid0,1,0,0]\leq 5$\\\cline{1-1}
$[-1,0,0,1,2\mid2,1,-1,0]\leq 4$ & $[1,1,1,0,0\mid0,0,0,1]\leq 5$\\\cline{2-2}
$[0,-1,0,1,2\mid1,2,-1,0]\leq 4$ & $[0,0,0,0,1\mid1,0,0,0]\leq 2$\\\cline{2-2}
$[0,1,2,-1,0\mid-1,0,1,2]\leq 4$ & $[0,0,1,1,1\mid1,1,0,0]\leq 5$\\\cline{1-2}
$[-1,0,-1,0,1\mid1,-1,0,0]\leq -1$ & $[-1,0,0,1,2\mid2,0,0,1]\leq 4$ \\\cline{2-2}
$[-1,-1,0,0,1\mid1,0,-1,0]\leq -1$ & $[-2,-1,0,1,2\mid2,-1,0,1]\leq 1$\\\cline{1-2}
$[1,0,0,0,0\mid 0,0,0,0]\le 2$     & $[0,0,0,0,-1\mid 0,0,0,0]\le -1$\\
\hline
\end{tabular}}
\end{center}

\begin{center}
\scriptsize{
\begin{tabular}{|c|c|}
\hline
\multicolumn{2}{|c|}{\sc \hspace{5mm} Spin-orbital Pauli constraints for high spin $d^8$ shell$\qquad$}  \\
\hline
$\quad[0,2,-1,0,1\mid1,-1,0]\leq 4\quad\,$ & $[-1,1,-1,0,0\mid1,-1,0]\leq -1$\\
$[0,2,0,1,-1\mid-1,0,1]\leq 4$ & $[-1,0,1,-1,0\mid0,1,-1]\leq -1$\\
$[0,1,2,-1,0\mid-1,1,0]\leq 4$ & $[-1,0,0,1,-1\mid1,-1,0]\leq -1$\\
$[0,1,2,0,-1\mid-1,0,1]\leq 4$ & $[-1,0,0,1,-1\mid0,1,-1]\leq -1$\\\cline{2-2}
$[0,0,1,2,-1\mid0,1,-1]\leq 4$ & $[0,0,0,-1,-1\mid-1,0,0]\leq -3$\\
$[0,0,1,2,-1\mid1,-1,0]\leq 4$ & $[0,0,-1,-1,0\mid0,0,-1]\leq -3$\\\cline{1-1}
$[0,1,-1,0,0\mid1,0,0]\leq 1$ & $[0,-1,0,0,-1\mid0,0,-1]\leq -3$\\
$[0,1,0,-1,0\mid0,1,0]\leq 1$ & $[0,0,-1,0,-1\mid0,-1,0]\leq -3$\\\cline{2-2}
$[0,1,0,0,-1\mid0,0,1]\leq 1$ & $[0,1,0,1,0\mid1,0,0]\leq 4$\\
$[0,0,1,0,-1\mid0,1,0]\leq 1$ & $[0,1,1,0,0\mid0,1,0]\leq 4$\\\cline{2-2}
$[0,0,0,1,-1\mid1,0,0]\leq 1$ & $[0,0,0,1,1\mid0,0,-1]\leq 3$\\\cline{1-2}
$[-1,1,0,0,1\mid1,-1,0]\leq 2$ & $[-1,0,0,0,-1\mid0,0,0]\leq -3$\\\cline{2-2}
$[-1,0,1,1,0\mid1,-1,0]\leq 2$ & $[-1,0,0,0,0\mid1,0,0]\leq -1$\\\cline{2-2}
$[-1,0,1,1,0\mid0,1,-1]\leq 2$ & $[1,0,0,0,0\mid0,0,0]\leq 2$\\\cline{2-2}
$[-1,0,1,0,1\mid1,0,-1]\leq 2$ & $[0,0,0,0,-1\mid0,0,0]\leq -1$\\
\hline
\end{tabular}}
\end{center}

Here  $[a_1,a_2,a_3,a_4,a_5\mid b_1,b_2,b_3]\le c$ is a shortcut for spin-orbital constraint
$$a_1\nu_1+a_2\nu_2+a_3\nu_3+a_4\nu_4+a_5\nu_5+b_1\mu_1+b_2\mu_2+b_3\mu_3\le c.$$
In the tables they are grouped into {\it cubicles\/}  with the same coefficients $a$, $b$, and $c$ up to an order within  $a$ and $b$.

The spin-orbital constraints for high spin configurations $d^3$ and $d^2$ can be obtained from $d^7$ and $d^8$ by  the {\it particle-hole duality\/}
$\nu_i\mapsto 2-\nu_{6-i}$ \cite{Klyachko08}. The term $d$-shell here refers to an orbital space of dimension 5 with no connotation to a specific physical interpretation.

In the remaining high spin configurations $d^N$,
$N=1,4,5,6,9$ the Pauli constraints
degenerate into equations
and can be described as follows.
\begin{itemize}
\item For a half-filled shell $N=5$ the multi-electron orbital space
$\wedge^5\mathcal{H} _\ell$ is one dimensional,
and the system effectively has only spin degrees of freedom.
Formally in this case $\mu_1=1,\nu_1=5$.
\item For $N=4$ spin and orbital occupation numbers are related by
the equation
$\mu_i=1-\nu_j, i+j=6$.
Similar equations $\mu_i=\nu_i-1$ hold for $N=6$.
\item For one electron system $N=1$ the Pauli constraints amount to
isospectrality $\mu_i=\nu_i$, while for
one hole ($N=9$) they take the form  $\mu_i=2-\nu_j, i+j=6$.
\end{itemize}
 A similar description holds for
orbital space $\mathcal{H}_\ell$ of any dimension $n$ containing
$1,n-1,n,n+1, 2n-1$ electrons in a high spin configuration. In particular, there are no nondegenerate spin-orbital constraints in $p$-shells.

Let's now
take a closer look at
some $d^7$  constraints
\begin{eqnarray}
\nu_1+\nu_{i+1}-3&\le&\mu_i\le1-(\nu_1-\nu_{i+1}),\quad i=1\ldots
4,\nonumber\\ \label{few}
M
&\ge&2(\nu_2+2\nu_3-\nu_4)-7.
\end{eqnarray}
The lower and the upper bounds on spin occupation numbers $\mu_i$ come from the second cubicles in the left and the right columns of the table respectively. Purely orbital constraints $\nu_1\le 2$, $\nu_5\ge1$ in the last line of the table ensure that the upper bound on $\mu_i$ is nonnegative.

It is worth to figure out what happens for $\nu_5<1$. Clearly, in this case the shell can't be in a high spin state, but it may be not in a low spin state either if the orbital constraint (\ref{low_spin1}) fails, i.e. $\nu_3+\nu_4-\nu_5>3$. Such a state with indefinite total spin  can not come from a spin independent interaction, like the Coulomb one. Thus we arrived at the following claim: {\it For any spin independent interaction in $d^7$ shell the following orbital  constraint holds
 \begin{equation}\label{spin_ind}
  \nu_3+\nu_4-\nu_5\le3.
 \end{equation}}
For example, no spin independent interaction can produce orbital occupancies like $(1.75,1.75,1.65.1.65,0.2)$.

The lower bound (\ref{few}) on the maximal magnetic moment $M=3\mu_1+\mu_2-\mu_3-3\mu_4$ is just the last constraint in the third cubicle on the left column. None of the constraints gives directly an upper bound on $M$ and one needs a suitable software, like \texttt{Convex} package \cite{FranzConv}, to extract this information:
\begin{align}
&M\le2(\nu_2+\nu_4)-3;
\framebox{$M\le 2(\nu_1+\nu_3-\nu_5)-1$}\;\framebox{Fe}\label{Fe_spin}\\ \nonumber
&M\le 9-2(2\nu_1-\nu_2+\nu_4);\;
M\le 9-2(\nu_2+2\nu_3-\nu_4);\\\label{Co_spin}
&\hspace{10mm}\framebox{$M\le 3\nu_3+(\nu_4-\nu_5)-5(\nu_1-\nu_2)$}\;\framebox{Co}\hspace{-15mm}
\end{align}
The framed constraints  set the actual Pauli bounds on spin magnetic moments for iron and cobalt for experimental orbital occupancies $\nu$.
Note also that for  spherically symmetric $d^7$-shell
with equal occupancies $\nu_i=\frac75$ the first marked  inequality gives
$M\le1.8\,\mu_B$. Any
excess
of this bound
must be accompanied
by
a non-spherical charge
deformation, which  many researchers
considered as a ``reminiscence of orbital magnetism"
 \cite{Brewer04}.
Pauli constraints provide another interpretation of this effect.


Spin populations in a spherical shell subject to the following
Pauli constraints
$$\textstyle\mu_1+\mu_2\le\frac45,\quad
\mu_1-\mu_3\le\frac25,\quad\mu_1\le2\mu_2+\mu_3,
$$
that confine
the maximal moment $1.8\,\mu_B$
to the unique spin configuration $\mu=(\frac35,\frac15,\frac15,0)$.
Surprisingly, the exclusion principle also produces an opposite
effect of {\it atomic spin polarization\/} induced by a given charge
density. An example of this provides the last inequality in
(\ref{few}) that set a {\it lower  bound\/}  on the magnetic moment
$M$.
To estimate the likelyhood
of this phenomenon,
let's
first
calculate constraints on the orbital occupancies compatible with
zero spin moment 
$$
\begin{array}{l}
\nu_1+\nu_3-\nu_4\le2,\quad\nu_1+\nu_4-\nu_5\le2,\quad\nu_2+\nu_3-\nu_5\le2,\\
\textstyle\nu_3-\nu_4-\nu_5\le\text{-}\frac34,\hspace{-4pt}\quad\nu_1+\nu_2\le\frac{13}{4},
\quad \quad\nu_2+2\nu_3-\nu_4\le\frac{7}{2}.
\end{array}
$$
Such $\nu$
constitute  a tiny fraction $\frac{1085}{31104}\sim3.5\%$
of the volume of all possible configurations,
i.e. a random electron density
does produce
a nonzero spin moment
with probability  $0.965$.
The ubiquity of this effect
raises the question whether $d^7$-shell can ever have a free spin
for some
orbital density $\nu$? As we have seen before,
the maximal moment $M=3\,\mu_B$
is bounded to the orbital configuraions
$\nu=(\nu_1,\nu_1,\nu_3,\nu_3,1)$, for which
the next to the last inequality in the above list
leaves
the narrow window $\frac{12}{8}\le
\nu_1\le\frac{13}{8}$.
A further calculation confirms that the spin population $\mu$ in
this interval is indeed free from constraints. An engineering of
such a state may be a
noble  endeavor, but the
general notion that atomic spins are free must be abandoned.
This never happens in high spin $d^7$ configuration neither in BCC nor in FCC
cubic field with orbital occupancies $(a,a,a,b,b)$ and
$(b,b,a,a,a)$ respectively. 
However, in low spin  $d^7$ configuration (\ref{low_spin1}) in FCC crystal field favoring  $e_g$ orbitals  $\nu_1=\nu_2=a\ge b=\nu_3=\nu_4=\nu_5$ there are no spin-orbital Pauli constraints except the initial one $\nu\le 2$. It is believed that a transition from low to high spin sector is responsible for the invar effect  in FCC Fe-Ni alloys.

In summary, the spin-orbital Pauli constraints contain a wealth of
an unexplored physical information. The above examples hardly
scratch the surface of it. They provide a quantitative version of the Pauli principle and depend neither on a force or on an interaction involved.

\paragraph{\bf Ferromagnetic iron.} The number of the Pauli constraints is drastically reduced in a
cubic solid where the orbital configuration is of the form
$\nu=(a,a,a,b,b)$, $a>b$ for a BCC crystal favoring $t_{2g}$
symmetry, and $\nu=(b,b,a,a,a)$, $b>a$ for an FCC structure.
For  $d^7$-shell
in a BCC crystal the spin-orbital Pauli constraints amount to 9
inequalities

{\vspace {-10pt}\footnotesize
\begin{eqnarray*}\label{spin-orb}
(3-2a)\mu_1+ (3-2a)\mu_2+ (2-2a)\mu_3+(2-2a)\mu_4\hspace{-2pt}&\le&\hspace{-2pt}0,\\
(2a-3)\mu_1+ (2a-3)\mu_2+(2a-4)\mu_3+ (2a-3)\mu_4\hspace{-2pt}&\le&\hspace{-2pt}0,\\
(11-7a)\mu_1+ (9-7a)\mu_2+(7-7a)\mu_3+
(9-7a)\mu_4\hspace{-2pt}&\le&\hspace{-2pt}0,\\
 (7a-11)\mu_1+
(7a-13)\mu_2+(7a-11)\mu_3+(7a-9)\mu_4\hspace{-2pt}&\le&\hspace{-2pt}0,\\
(a-1)\mu_1+(a-3)\mu_2+(a-2)\mu_3+(a-2)\mu_4\hspace{-2pt}&\le&\hspace{-2pt}0,\\
(2-a)\mu_1 -a\mu_2+ (1-a)\mu_3
-a\mu_4\hspace{-2pt}&\le&\hspace{-2pt}0,\\
(3a-4)\mu_1+(3a-6)\mu_2 +(3a-5)\mu_3+(3a-6)\mu_4\hspace{-2pt}&\le&\hspace{-2pt}0,\\
(9a-17)\mu_1 +(9a-15)\mu_2 +(9a-13)\mu_3 +(9a-11)\mu_4\hspace{-2pt}&\le&\hspace{-2pt}0,\\
(23-15a)\mu_1+(17-15a)\mu_2+(19-15a)\mu_3+(21-15a)\mu_4\hspace{-2pt}&\le&\hspace{-2pt}0,
\end{eqnarray*}}
\hspace{-5pt} depending on  the occupation number $a$  of a $t_{2g}$
orbital.
Below I will
focus on the
archetypical example of
BCC iron whose magnetic moment primary comes from  spins of
partially filled $3d$-shells.
Its electronic density was accurately measured by Jauch\&Reehuis \cite{Jauch07}. They also addressed the question whether $3d$ electron are indeed localised or itinerate by  testing the experimental data versus three available band theory calculations. All of them predict an  expansion of the atomic $d$-shell in a solid,  
while the experiment shows a clear  contraction. The authors concluded: {\it no indication for a failure of the localized $3d$ electron model is noticed} \cite{Jauch07}.
The authors also found that the crystal field changes the free atomic configuration $d^6$ into
$d^7$, and  $t_{2g}$ electrons make up $62.5(3)\%=5/8$ of the
$d$-shell. Whence $a=35/24\approx1.458(7)$.

Since the spin density matrix
commutes with the {\it magnetic\/} symmetry group,
the states $|\frac32\rangle$, $|\frac12\rangle$,
$|\text{-}\frac12\rangle$, $|\text{-}\frac32\rangle$ with definite
spin projections onto the magnetization axis are its eigenvectors.
The natural spin occupation numbers $\mu=(\mu_1,\mu_2, \mu_3,
\mu_4)$ are just the probabilities to find an iron atom in one of
these states.

The Pauli constraints on the magnetic moment
$M=3\mu_1+\mu_2-\mu_3-3\mu_4$
are shown graphically in Fig.~1. Observe, first
of all, that the
iron magnetic
moment
is the maximal possible for the
given electron distribution.
This suggests
that the spins within the Pauli constraints
are indeed free. It worth also to note that
in the
segment $AB$
the saturation magnetic moment and the respective spin distribution
are defined by the parameter $a$
$$
\textstyle M_\text{sat}=7a-8,\quad
\mu_{\text{sat}}=\left[\frac32(a-1),\frac12(a-1),3-2a,0\right].
$$
The vertex $A$ corresponds to the spherical atom. The maximal
moment in a BCC crystal can not exceed
$2.5\,\mu_B$ attained at the vertex $B$ with
$\mu=\left(\frac34,\frac14,0,0\right)$. This is the only point where
the atom is always in an up-spin state. In the segment $BC$ the
saturation moment and spin occupations are as follows
$$\textstyle M_\text{sat}=16-9a,\quad
\mu_{\text{sat}}=\left[\frac12(9 -5a),\frac12(a-1), 2a-3,0
\right].$$ The point $C$ lies on
the boundary between high-
and low-spin configurations.
It is surprising
that at the high- to low-spin  transition
the ordered magnetic moment
retains its value $M=1\,\mu_B$. Finally, the parameter
$a=\frac{14}{9}$ at the point $D$ marks the onset of strictly positive
magnetic moment.

\paragraph{\bf Cobalt and nickel spin moments.}
Below we calculate Pauli bounds on spin moment for cobalt and nickel and compare them with the following   experimental data 
deduced from gyromagnetic ratios
\cite{Stearns84} and the total moments \cite{Pauthenet82}
\begin{align*}\label{FeCoNi_Spin&Orb}
M^\text{Fe}_\text{tot}&=2.226\,\mu_B;&
M^\text{Fe}_\text{spin}&=2.143\,\mu_B;&
M^\text{Fe}_\text{orb}&=0.083\,\mu_B;\\
M^\text{Co}_\text{tot}&=1.729\,\mu_B;&
M^\text{Co}_\text{spin}&=1.577\,\mu_B;&
M^\text{Co}_\text{orb}&=0.152\,\mu_B;\\
M^\text{Ni}_\text{tot}&=0.619\,\mu_B;&
M^\text{Ni}_\text{spin}&=0.563\,\mu_B;&
M^\text{Ni}_\text{orb}&=0.056\,\mu_B.
\end{align*}

At low temperature cobalt has hexagonal crystal structure that splits $d^7$-shell into three sectors
$$
a_g=\langle 3z^2-r^2\rangle,\quad e_g=\langle xz,yz\rangle,\quad e_g^\prime=\langle xy,x^2-y^2\rangle
$$
 spanned by eigenvectors of the orbital moment operator $L_z=i(x\partial_y-y\partial_x)$: $a_g=|0\rangle$, $e_g=|\pm1\rangle$, $e_g^\prime=|\pm2\rangle$.  
Here are their experimental orbital populations $(a,b,b,c,c)=(\nu_1,\nu_2,\nu_3,\nu_4,\nu_5)$
found by Jauch\&Reehuis \cite{Jauch09}
\begin{equation*}\label{Co_occ2}
a=1.846(22);\quad b=1.2935(80);\quad
c=1.2835(95).
\end{equation*}
They set  the Pauli bound (\ref{Co_spin}) on spin magnetic moment
\begin{align}\label{Co_SpinBnd1}
M^{\mathrm{Co}}_\text{spin}&\le
3\nu_3+(\nu_4-\nu_5)-5(\nu_1-\nu_2)\\
&=8b-5a=1.118\,\mu_B,\nonumber
\end{align}
which is clearly incompatible with the experimental  value $M^{\mathrm{Co}}_\text{spin}=1.577\,\mu_B$.
The
discrepancy came from the fact that Jauch\&Reehuis measured {\it populations} of the subshells, while the Pauli constraints refer to {\it eigenvalues} of the density matrix.      
In absence of the orbital moment 
they would be the same, but otherwise  we have to take it into account
by a variation of the orbital eigenstates'  occupancies
$
(a,b,b,c,c)\mapsto (a,b+\varepsilon,b-\varepsilon,c+\delta,c-\delta)
$
that preserves the number of electrons in the subshells $|0\rangle$, $|\pm1\rangle$, $|\pm2\rangle$   and matches  the experimental value of the orbital moment $M_\text{orb}^\text{Co}=2\varepsilon+4\delta=0.152\mu_B$.

The Pauli bound (\ref{Co_SpinBnd1}) depends on the exact order of the occupancies $a,b\pm\varepsilon, c\pm\delta$. Observe that within the error bar $b=c$ and therefore  the order is determined by the type of inequality $\varepsilon \lessgtr\delta$. For $\varepsilon<\delta$ we have  $$(\nu_1,\nu_2,\nu_3,\nu_4,\nu_5)=(a,c+\delta,b+\varepsilon,b-\varepsilon,c-\delta)$$
that still gives  too low Pauli bound
\begin{align*}
M_\text{spin}^\text{Co}&\le3\nu_3+(\nu_4-\nu_5)-5(\nu_1-\nu_2)\\ &=3(b+\varepsilon)+(b-\varepsilon)-(c-\delta)-5(a-c-\delta)\\
&=7(2-a)+2\varepsilon+6\delta=1.078+M_\text{orb}+2\delta\\
&\textstyle\le 1.078+\frac32 M_\text{orb}=1.306\,\mu_B.
\end{align*}
For $\varepsilon>\delta$ we have to switch $b\leftrightarrows c$ and $\varepsilon\leftrightarrows\delta$ that gives
$$M_\text{spin}^\text{Co}\le 7(2-a)+6\varepsilon+2\delta.$$
When $\delta=0$, i.e. the orbital moment comes entirely from $e_g$ subshell,  we get the maxmal Pauli bound
$$M_\text{spin}^\text{Co}\le 7(2-a)+6\varepsilon=7(2-a)+3M^\text{Co}_\text{orb}=1.534\mu_B,$$
which is slightly less than
the experimental value $1.577\,\mu_B$ but indistinguishable from it
in view of
the orbital occupancy uncertainty $a = 1.846(22)$.

In summary, there is unique way to reconcile the
orbital moment of cobalt with Pauli constraints.
The moment comes from $e_g$ subshell
and has minimal possible value
compatible with the Pauli bound. The latter, in turn, coincides with the observed spin moment.
In a sense, the ferromagnetic order parameter    generates the orbital moment
to raise  the Pauli bound by about 40\%. That kind of behavior can be expected  in the elemental ferromagnet with strongest known magnetic interaction
as seen in its  high Curie temperature.




For nickel, which is the weakest elemental ferromagnet, the above analysis based on the Pauli constraints is inconclusive.
There are four Pauli bounds on spin magnetic moment $M=2\mu_1-2\mu_3$ for high spin  $d^8$ configuration
\begin{align*}
M&\le
 \nu_1-\nu_2+\nu_3-\nu_4+\nu_5,&
 M&\le 2\nu_1-2\nu_3-2\nu_5+4,\\
 M&\le 2\nu_2-2\nu_4+4\nu_5-4,&
 M&\le 4\nu_3+2\nu_4-2\nu_2-4,
\end{align*}
where the orbital occupancies are subject to additional constraints
$\nu_1\le2$, $\nu_1+\nu_5\ge3$. When the latter one fails the shell collapses   into a nonmagnetic spin $S=0$ state. Such collapse can't happen in FCC crystal with orbital occupancies $(a,a,b,b,b)$ favoring $e_g$ orbitals but may  occur  in BCC or tetragonal environment. Apparently this happens in Ni alloy with about 12 atomic percents of vanadium, so that in average a Ni atom has about 1.5 impurities among its 12 nearest neighbors that may reverse the inequality $\nu_1+\nu_5\ge3$.

For spherically symmetric shell with equal occupancies $\nu_i=\frac85$ the second bound gives $M\le\frac45=0.8\mu_B$. None of the first three constraints can produce a smaller bound for no orbital occupancies. Therefore nickel spin moment  $M_\text{spin}^\text{Ni}= 0.563\mu_B$ can saturate, if any, only the last constraint.
However, the orbital occupancies $\nu=(a,a,b,b,b)$ of $e_g$ and $t_{2g}$ subshells found in \cite{Jauch08}
\begin{align*}\textstyle n_e= 2a= 3.475(16),\qquad 
n_t=3b=4.525(16),
\end{align*}
give the lowest Pauli bound $M\le1.442\mu_B$, coming from the first constraint, which is
not even close to 
the actual value. 
For cobalt we have found unique way to comply with the Pauli constraints, while for nickel they hold from the beginning and any change without additional experimental data  would be purely speculative. There is no reason to believe that 
for a weak ferromagnetic interaction the Pauli spin bound must be saturated.




\paragraph{\bf Pauli constraints and statistics.}
The next example deals with an impact of the Pauli constraints on
spin statistics. To begin with, consider a system of
noninteracting free spins in a lattice at temperature $T$ and
magnetic field $H$. Its thermal properties are governed by
Helmholtz free energy per site
\begin{equation}\label{Helmh}F(\mu)=-Hm\cdot\mu-k_BT S(\mu), \end{equation}
where $m$ and $\mu$ are vectors of possible local quantized
magnetic moments $m_i$ and the respective probabilities $\mu_i$,
$S(\mu)=-\sum\mu_i\log\mu_i$ is the magnetic entropy. The thermal
equilibrium corresponds to the minimum of $F(\mu)$ given by the
Gibbs canonical distribution
\begin{equation}\textstyle\label{regime1}\mu_i\sim e^{m_i\beta},
\quad\beta=H/k_BT.
\end{equation}

In real solids the orbital density matrix of an atom is fixed by the
Coulomb crystal field. This in turn  imposes Pauli constraints on
the probabilities $\mu_i$ to find an atom in a given spin state. As
a result, the thermal equilibrium is attained at a {\it relative
minimum\/} of the free energy taken over the {admissible\/}
probabilities $\mu$ that form the so called {\it moment polytope\/}.
Its boundary separates available spin configurations from
forbidden  ones and plays a role similar to the Fermi surface.
Fig. 2 shows it for BCC iron.
\begin{figure}[h]
\includegraphics[width=3.5cm]{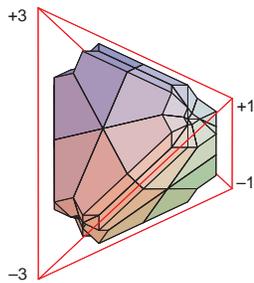}
\caption{\label{fig2} The moment polytope for iron inside the
tetrahedron with barycentric coordinates $\mu$. The numbers
indicate magnetic moments at vertices of the tetrahedron. }
\end{figure}

The actual thermal evolution in BCC configuration $d^7$ is rather
complicated and strongly depends on the orbital occupation number
$a$. The experimental value $a=\frac{35}{24}$ for iron falls into
the interval $\left[\frac75,\frac{19}{13}\right]$ where the
evolution can be described as follows.

It begins with  the uniform distribution at $\beta=0$ and follows
Gibbs law (\ref{regime1})
until $\mu$ saturates {\it the third\/} Pauli constraint in the
list of 9 constraints on page
\pageref{spin-orb}, turning it into equation which we write in
the symbolic form
\begin{equation}\label{Facet_1}
 c_1\mu_1+c_2\mu_2+c_3\mu_3+c_4\mu_4=0.
\end{equation}
This happens at the critical value
\begin{equation}\label{crit1}\textstyle\beta_1=\frac12\log\alpha,\quad
 c_1\alpha^3+c_2\alpha^2+c_3\alpha+c_4=0,\quad\alpha>1.
\end{equation}
Starting from this point the equilibrium state evolves along the
saturated facet (\ref{Facet_1}) with the  Gibbs distribution
replaced by
\begin{equation}\label{regime2}
\textstyle\mu_i\sim e^{m_i\beta+\gamma c_i},
\end{equation} where
the Lagrange multiplier $\gamma$ is determined by equation of the
facet (\ref{Facet_1})
$$\textstyle
\sum_i c_i e^{m_i\beta+\gamma c_i}=0.
$$
This form persists until the second critical point
\begin{equation}\textstyle\label{crit2}\beta_2=\frac14\log\frac{20-12a}{19-13a}
\end{equation}
where   the thermal trajectory hits  {\it the first\/} Pauli
constraint in the same list turning it into equation
\begin{equation}\label{Facet_2}
d_1\mu_1+d_2\mu_2+d_3\mu_3+d_4\mu_4=0.
\end{equation}
From this point on the thermal equilibrium
\begin{equation}\label{regime3}\mu_i\sim e^{m_i\beta+\gamma c_i+\delta d_i}
\end{equation} evolves along
the intersection of two facets (\ref{Facet_1}) and
(\ref{Facet_2}). The parameters $\gamma,\delta$ are fixed by the
facets' equations
$$\textstyle \sum_i c_ie^{m_i\beta+\gamma c_i+\delta d_i}=0,\qquad
\sum_i d_ie^{m_i\beta+\gamma c_i+\delta d_i}=0. $$
As $\beta\rightarrow\infty$ the equilibrium
eventually freezes  up at the vertex of the moment polytope
\begin{equation}\label{mu_sat}\textstyle\mu_{\text{sat}}=\left[\frac{3}{2}(a-1), \frac{1}{2}(a-1),
3-2a,0\right]
\end{equation}
with the maximal magnetic moment $M_{\text{sat}}=7a-8$.

In real ferromagnets, the internal field $H$ entering in the
parameter $\beta=H/k_BT$ is an unknown function of the local
magnetic moment $M=3\mu_1+\mu_2-\mu_3-3\mu_4$.
In a way of illustration assume, following Weiss, that $H$ is
proportional to $M$. Figure 3 shows the resulting magnetization
curve.
\begin{figure}[t]
\includegraphics[width=6cm]{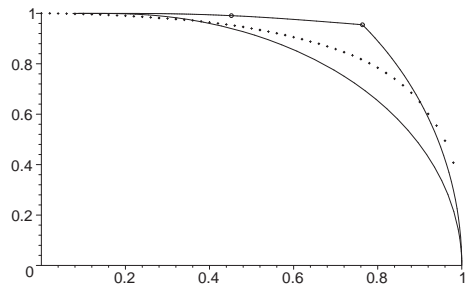}
\caption{
\label{fig3} Reduced magnetization
$M/M_{\text{sat}}$ versus reduced temperature $T/T_C$ for Weiss
spin-3/2 model, its modification
by the  Pauli constraints discussed in the text, and the
experimental points for iron \cite{Crangle71}. Circles mark the
critical points. }
\end{figure}
Clearly, the spin-orbital Pauli constraints change the temperature
behavior both quantitatively and qualitatively. The latter
modification,
epitomized in the critical points, could be seen in the actual
thermal evolution as crossovers between the different physical
regimes (\ref{regime1}), (\ref{regime2}), (\ref{regime3}).

The experimental value $a=\frac{35}{24}$  gives critical parameters
$\beta_1=0.55429$, $\beta_2=1.02359$ and the respective magnetic
moments $M_1/M_{\text{sat}}=0.95585$, $M_2/M_{\text{sat}}=0.99296$.
By interpolation of the magnetization data \cite{Crangle71} the
latter can be converted into critical temperatures
$T_1=461^\circ\,\text{K}$ and $T_2=176^\circ\,\text{K}$ where the
crossovers are expected to happen. Above $T_1$ spins detach  from a
direct influence of the lattice and follows usual Weiss theory with
no additional Pauli constrains. Therefore it may be not surprising
that $T_1$ essentially coincides with Debye temperature of iron. The
numerical value of
$T_2$ is very sensitive to small variations of $a$ and should be
considered as a rough estimation.

The first critical point can be identified with a crossover at
$467^\circ\,\text{K}$ previously reported by K\"obler
\cite{Koebler03},
At the second one,
the angle between two adjacent segments of the modified Weiss curve
in Fig.~3 is too small to see the crossover directly in the
magnetization $M(T)$. Therefore we look for the effect in its
derivative $\chi_{\hspace{0pt}_{HF}}(T)=\frac{\partial
M(T)}{\partial H_{\text{ext}}}$ measured at an external field
$H_{\text{ext}}$ high enough to make iron crystal a single magnetic
domain. The experimental data \cite{Pauthenet82} are shown in Fig.~4
together with a quadratic component that may come from the Bloch
correction to Pauli paramagnetism \cite{Bloch29} or any other source
unaffected by the Pauli constraints. The residue shows a sharp
crossover around $200^\circ\,\text{K}$.

Decreasing the reported orbital occupation number $a=\frac{35}{24}$
by negligible quantity 0.00085, well within the error bar
$\pm0.007$, gives the critical temperatures
$T_1=465^\circ\,\text{K}$, $T_2=200^\circ\,\text{K}$ compatible with
both observations.
\begin{figure}[t]
\includegraphics[width=5.5cm]{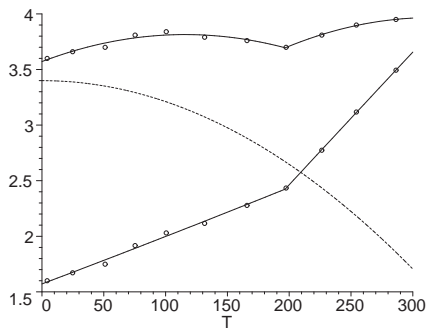}
\caption{\label{fig4} {
High field susceptibility $\chi_{\hspace{0pt}_{HF}}$ for iron \cite{Pauthenet82} in
units $10^{-6}\text{cm}^3/g$ versus temperature in $K^\circ$ (the upper
curve), its quadratic component $aT^2+b$ (shown in dots), and the
residue representing
the pure quantum crossover effect.
} \vspace{-0.3cm}}
\end{figure}

Crossover phenomena are widely spread in magnetic compounds. They
have been collected and classified in recent book \cite{Koebler09}
which may well contain more examples of physical effects caused by
the spin-orbital Pauli constraints. Among possible reasons for the
crossover phenomena  the authors mentioned a change in the number
of relevant spin states. This sounds as a paraphrase of the
description given above: crossing a critical point turns a Pauli
inequality into the equation that effectively reduces available
spin degrees of freedom.

\paragraph{\bf Conclusion.}
This study rests upon a novel concept of spin-orbital Pauli constraints that provide a quantitative version of the Pauli principle.
Being universal and ubiquitous they may have many applications ranging from chemistry to nuclear physics.   This paper, however,
is focused on ferromagnetism as the most striking and straightforward  macroscopic manifestation of spin.

Unfortunately till now there is no rigorous theory of ferromagnetism, mainly because in non-relativistic quantum mechanics spins are not directly involved  in the interaction and respond only to a still not fully understood force-free Pauli magic.
The latter
challenges monopoly of
Newton's {\it culture of force} \cite{Wilczek04}, but due to its intractability
is often
substituted
by a spin dependent
{\it exchange interaction\/}. For those who take it soundly
I recall the celebrated Lieb-Mattis theorem \cite{LiebMattis,Peierls79}: {\it For a system of electrons confined to a line and subject to an arbitrary symmetric  spin independent interaction the ground state can't be magnetic.} At the same time the exchange approximation applied to a suitable system of that type gives rise to the Heisenberg model with a magnetic ground state. A ghost of this artefact will follow us forever and may show up any time the exchange interaction is invoked.

It should be also emphasized that
spin independent hamiltonian commutes with the total spin $S$ and makes it a well defined quantum number.
Therefore a precursor of ferromagnetism must be a high total spin sector, from which a ferromagnetic state may emerge by breaking its spin degeneracy. Any surrogate spin dependent Hamiltonian
would kill this fundamental truth. This scenario essentially calls for a version of Hund's rule that determines the  total spin state of a multi-electron system. As we have seen above admissible electron densities depend on the total spin. For example, $d^7$ shell must be in a low spin state if $\nu_5<1$ and $d^8$ shell collapses into a spinless   state for $\nu_1+\nu_5<3$.


Anyway, the
spin-orbital Pauli constraints provide a unique known mechanism that affects noninteracting spins.
Therefore they must play a crucial role in formation of a ferromagnetic state.
In many cases the natural orbital occupancies $\nu_i$ determine the total spin sector.

To support this claim, the Pauli constraints were calculated for all high spin $d$-shell configurations
and tested in ferromagnetic iron, and to a smaller extend in cobalt and nickel.
It turns out that for iron and cobalt, where the ferromagnetic interaction is strong
and epitomised in high  Curie temperature, the spin magnetic moments attain maximal values
allowed by the Pauli constraints. This may be the first experimental support of relevance the Pauli constraints. This result  also resolves  80 years old Pauli problem on the origin of reduction the atomic magnetic moments in ferromagnets.
For
nickel where the ferromagnetic interaction is much smaller the actual moment is below the Pauli bound.

We are gratefully acknowledge W.~Pauli  for a valuable insight.

\end{document}